\documentclass[twocolumn,trackchanges]{aastex62}

\usepackage{comment}
\usepackage{amsmath}
\usepackage[utf8]{inputenc}
\usepackage{layouts}
\usepackage{amsmath}
\usepackage{hyperref}

\graphicspath{{./}{figures/}}

\received{\today}
\revised{XXX}
\accepted{XXX}
\submitjournal{ApJ}

\shorttitle{Chromospheric wave heating with IRIS and Bifrost}
\shortauthors{Molnar et al.}

\begin{document}

\title{Constraining the systematics of (acoustic) wave heating estimates in the solar 
chromosphere}

\correspondingauthor{Momchil Molnar}
\email{mmolnar@ucar.edu}

\author[0000-0003-0583-0516]{Momchil E. Molnar}
\altaffiliation{Currently at the High Altitude Observatory, NCAR.}
\affil{National Solar Observatory, Boulder, Colorado, USA}
\affil{Department of Astrophysical and Planetary Sciences, University of Colorado, Boulder, USA}
\affil{Laboratory for Atmospheric and Space Physics, University of Colorado, Boulder, USA}

\author[0000-0001-8016-0001]{Kevin P. Reardon} 
\affil{National Solar Observatory, Boulder, Colorado, USA}
\affil{Department of Astrophysical and Planetary Sciences, University of Colorado,
	Boulder, USA}

\author[0000-0002-3699-3134]{Steven R. Cranmer} 
\affil{Department of Astrophysical and Planetary Sciences, University of Colorado, Boulder, USA}
\affil{Laboratory for Atmospheric and Space Physics, University of Colorado, Boulder, USA}

\author[0000-0001-7458-1176]{Adam F. Kowalski} 
\affil{National Solar Observatory, Boulder, Colorado, USA}
\affil{Department of Astrophysical and Planetary Sciences, University of Colorado, Boulder, USA}
\affil{Laboratory for Atmospheric and Space Physics, University of Colorado, Boulder, USA}

\author[0000-0002-0189-5550]{Ivan Mili\'c}
\affil{Astronomical Observatory, Belgrade, Serbia}
\affil{National Solar Observatory, Boulder, Colorado, USA}



\begin{abstract}

Acoustic wave heating is believed to contribute significantly to the missing energy input 
required to maintain the solar chromosphere in its observed state. 
We studied the propagation of waves above the acoustic cutoff
in the upper photosphere into the
chromosphere with ultraviolet and optical spectral
observations interpreted through comparison with three dimensional
radiative magnetohydrodynamic (rMHD) \emph{Bifrost} models to constrain the 
heating contribution from acoustic waves in the solar atmosphere. 
Sit-and-stare observations taken with
the Interface Region Imaging Spectrograph 
(IRIS) and data from the 
Interferometric BIdimensional Spectrograph (IBIS) 
were used to provide the observational basis of this work.
We compared the observations with synthetic observables derived from
the Bifrost solar atmospheric model. 
Our analysis of the \emph{Bifrost} simulations show that 
internetwork and enhanced network regions exhibit significantly different wave 
propagation properties, which are important for the accurate wave flux estimates.
The inferred wave energy fluxes based on our observations
are not sufficient to maintain the solar 
chromosphere. We point out that the systematics of the
modeling approaches in the literature lead to differences which could
determine the conclusions of this type of studies, based on the same observations.

\end{abstract}

\section{Introduction}
\label{sec:Introduction}

The solar chromosphere has a higher temperature than expected from 
radiative equilibrium
\citep{1977ARA&A..15..363W, 2019ARA&A..57..189C}. 
The additional heating required to maintain the 
chromosphere in its observed thermodynamic state 
is approximately a few to tens of kW/m$^2$, depending on the 
activity of the particular solar feature
\citep{1976ASSL...53.....A, 2021A&A...647A.188D}. Understanding
the primary heating sources is important for modeling the solar
chromosphere correctly, as these will determine 
its structure and observed properties. 
This is an important astrophysical question beyond the Sun, because stellar
chromospheres are the source of the UV continuum that influences their
surrounding environment, e.g. dictates the atmospheric chemical composition 
of their exoplanets \citep{2017ARA&A..55..159L}.

Previous work has suggested that the
two most viable mechanisms to provide 
the missing heating in the solar atmosphere is through 
stochastic release of stored magnetic energy or dissipation of 
magnetohydrodynamic (MHD) waves. 
Release of magnetic energy -- either through current sheet dissipation
\citep{2005ApJ...633L..57S, 2021A&A...652L...4L}
or magnetic reconnection~\citep{1997Natur.386..811I, 2017ApJ...851L...6R} has been 
reported throughout the chromosphere with limited global 
heating implications.
Conclusive observational evidence of this heating
process is still lacking, 
even if models predict it to be pervasive in the active Sun 
\citep{2022A&A...661A..59D}.

In this paper we focus on the other possible heating mechanism 
-- acoustic wave energy 
dissipation. Chromospheric heating by waves was proposed
in the late 1940's 
\citep{1946NW.....33..118B, 1949AnAp...12..203S}
and has been discussed extensively in the literature 
\citep[see][for a short review]{2019ASSL..458.....A}. 
Recent progress on constraining the 
wave heating in the solar chromosphere has been
enabled by the technological advances
of adaptive optics, tunable filtergraphs and 
more sensitive ultraviolet (UV) and near-infrared (IR) instruments. 
There are two differing conclusions 
about the energetic significance of acoustic waves
in the lower solar atmosphere. In general, the body of work
based on high-cadence Doppler velocity observations 
interpreted with a 1D static atmospheric perturbative approach
derive wave fluxes sufficient to maintain the quiet chromosphere
\cite[e.g.][]{2009A&A...508..941B,2016ApJ...826...49S, 
2020A&A...642A..52A}. On
the other hand, studies based on Doppler velocities from and UV/mm continuum
observations interpreted with 1D time-dependent radiative hydrodynamic (rHD)
models,
well suited for chromospheric studies, come to the opposite conclusion -- 
acoustic waves do not carry sufficient energy flux to maintain the quiet 
chromosphere \citep{2005Natur.435..919F, 2007PASJ...59S.663C, Molnar_2021}.
However, the latter studies have been critiqued for systematic biases toward 
underestimating the acoustic flux \citep{2007ASPC..368...93W}.

For this project, we extended the previous work on determining the 
acoustic wave flux in the chromosphere with optical
observations of \cite{Molnar_2021} (henceforth Paper I) with UV data
of the low and high chromosphere from the Interface Region Imaging 
Spectrograph \citep[IRIS, ][]{2014SoPh..289.2733D}. We also used
3D instead of 1D radiative MHD (rMHD) models to interpret the wave observations. 
This could be considered an extension 
of the similar work by \cite{2021A&A...648A..28A} with the inclusions of
multiple  spectral lines in the IRIS UV spectral sampling interval,
instead of relying on the wings of the \ion{Mg}{2} h and k lines.
We argue that the interpretation of
the observed oscillatory signals requires the use of 3D MHD models,
contrary to 1D models used in previous work. 
Wave modeling that relies on 1D semi-empirical models 
\citep[such as those of][]{2011JGRD..11620108F} calculate the 
properties of the observed waves as perturbations on a static atmosphere,
which may be an inaccurate approximation, if the dynamical 
oscillations are maintaining the atmosphere
in a dynamic state far from equilibrium \citep[see][for a treatment of the
similar physical setting
on Mira-like stars]{Bertschinger_Chevalier_1985}.

This paper describes the observed wave properties
in the lower and upper chromosphere 
observed in the UV with IRIS and tries to infer 
the energy flux of acoustic waves propagating in these regions 
through comparison with spectral synthesis from 3D rMHD 
Bifrost models \citep{2011A&A...531A.154G}. We compared
those results with diagnostics from the optical part of the spectrum
obtained with the Imaging BIdimensional Spectrograph 
\citep[IBIS,][]{2006SoPh..236..415C}.
The paper is organized in the following way: 
Section \ref{sec:Observations} describes 
the UV and optical observations used throughout the paper;
Section \ref{sec:power_spectra} presents
the derived properties of the power spectra of different 
diagnostics;
Section~\ref{sec:Modeling} presents the wave diagnostics derived
from synthetic observables from \textit{Bifrost} MHD enhanced network 
models; Section~\ref{sec:Uncertainties} discusses the systematics between different 
modeling approaches. We conclude with the wave-energy flux estimates in 
Section~\ref{sec:Inferred_flux} and discuss the implications our results in 
Section~\ref{sec:Conclusions}.

\section{Observations}
\label{sec:Observations}

To extend the previous work in Paper I, 
we use UV spectral diagnostics 
observed with the IRIS spacecraft to sample the upper chromospheric velocity 
and intensity diagnostics.
For this paper, we concentrate in this paper on the 
\ion{Mn}{1} 280.108 nm 
line~\citep[lower chromosphere, ][]{2013ApJ...778..143P},
the \ion{Mg}{2} h$_{2}$ \& k$_{2}$ features (middle chromosphere) and
the \ion{Mg}{2} h$_{3}$ \& k$_{3}$ features
\citep[upper chromosphere, ][]{2013ApJ...772...90L}. The IRIS data 
archive offers a vast collection of observations 
containing this spectral line set.
We compare the properties of the UV data with the results
from Paper I to obtain new and more complete estimate for the 
energy fluxes that acoustic waves are carrying
and the possible chromospheric heating implications.

\begin{figure*}
	\includegraphics[width=0.95\textwidth]{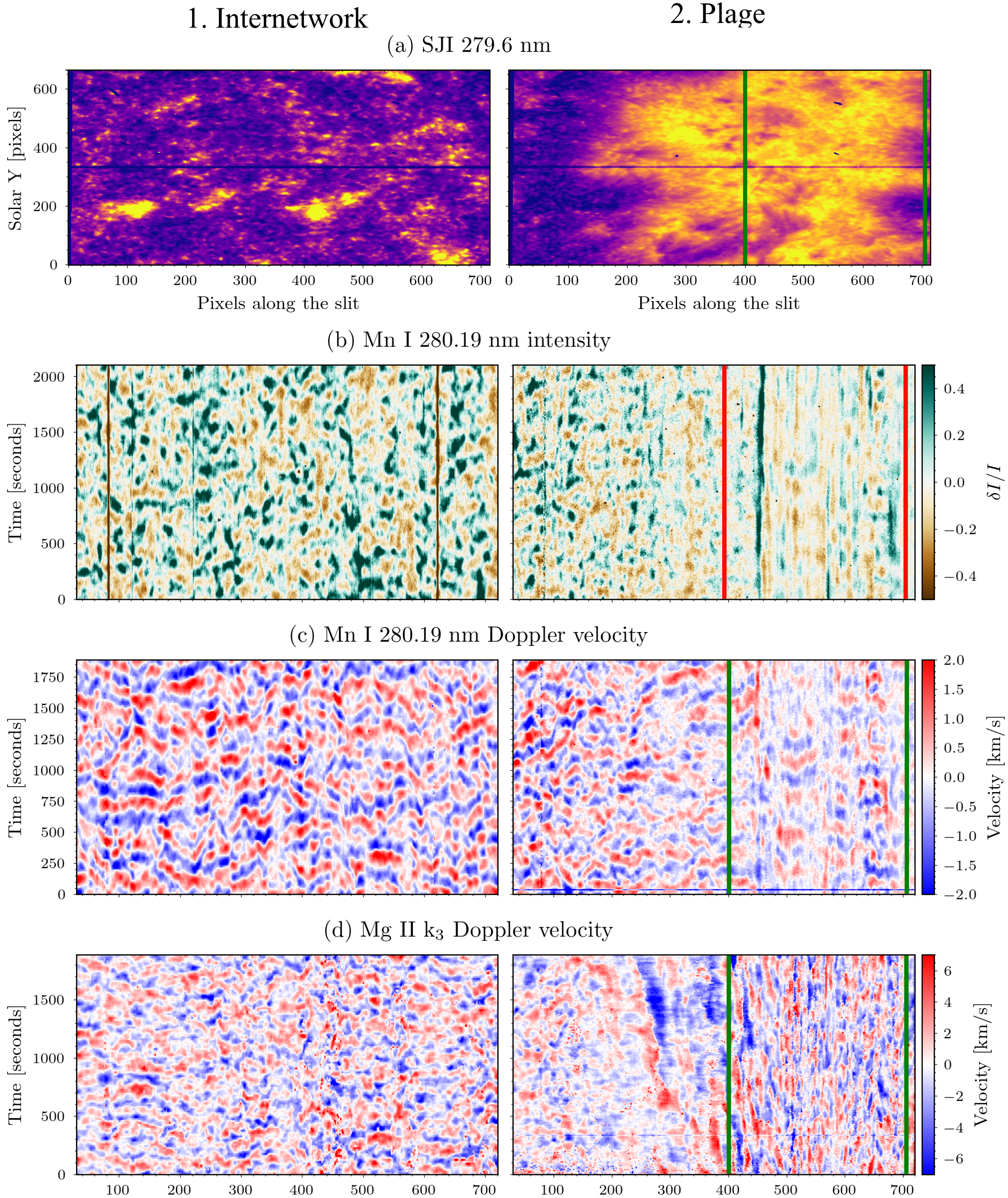}
	\caption{The data used in this study 
		comes from two different regions -- internetwork
		(left column) and plage (right column), where the dark line across the center of the image is the actual slit.
		The left column shows observations of an internetwork region from 
		2013 November 16; the right column of a plage region observations
		from 2014 September 18 (see Table~\ref{tab:Observations_IRIS}).
		The top row (a) are slit-jaw images 
		in the 279.6 nm spectral window for the internetwork (left) and plage (right).
		Row (b) show the 
		relative intensity variations (to the mean intensity at the 
		particular slit position)
		at the core of the \ion{Mn}{1} 280.19 nm line. Rows (c) and (d) 
		show the Doppler velocities derived from the \ion{Mn}{1} 280.19 nm line
		and the \ion{Mg}{2} k$_3$ feature, respectively. All panels present along their 
        x-axis the slit dimension.}
	\label{fig:diagnostics_overview}
\end{figure*}

Throughout the paper we will discuss two different types of solar features:
\emph{internetwork} and \emph{plage}. These regions harbor
weak magnetic fields in the case of internetwork and stronger
magnetic fields in the case of the plage.
The choice of these two types of solar surface is based on 
their relative simple discrimination from the rest of the solar structures. 
Furthermore, in the internetwork we didn't expect the weak magnetic 
field to be significant for the wave propagation. In the case of the plage, 
previous work showed the ubiquity of fluctuation signatures and a mostly 
vertical magnetic field \citep{2020A&A...644A..43P,2021ApJ...921...39A}, which 
could harbor MHD wave modes. 

\subsection{Processing of the IRIS data}
\label{subsec:IRIS_data_processing}

\begin{table}[!ht]
	\begin{center}
		
		\label{tab:Observations_IRIS}
		
		\begin{tabular}{c|c|c|c|c} 
			Date & Start [UT] & End [UT] &
			Cadence [s] & Solar feature \\ \hline
			20131116 & 07:33 & 08:08 &17.0 & Internetwork \\
			20140918 & 10:19 & 12:16 & 9.4  & Plage \\
		\end{tabular}
		\caption{IRIS observations used in this work.}
	\end{center}
\end{table}

We use the \textit{level\_2} spectral rasters from the IRIS online data
archive\footnote{https://iris.lmsal.com/data.html}
for this analysis. The particular datasets used in this study are described in 
Table~\ref{tab:Observations_IRIS}. We chose two sets of observations
from the earlier stages of the 
IRIS mission to ensure higher sensitivity and lower noise levels.
The datasets used in this work are in sit-and-stare mode, 
which increases the signal-to-noise ratio of the observations and provides
higher Nyquist sampling frequency. 

The two UV spectral lines of interest have different shapes 
-- the \ion{Mn}{1} 280.1 nm line has a simple absorption profile, whereas 
the \ion{Mg}{2} h \& k lines have a complicated, typically 
double-peaked shape due to 
the high opacity non-equilibrium effects at chromospheric heights
\citep{1967ApJ...149..239T}.
We adopted different 
fitting approaches to extract the physical parameters from the two spectral lines.
The \ion{Mg}{2} h \& k lines are fitted with the IDL 
routine \emph{iris\_get\_mg\_features\_lev2.pro}, part of the 
SSW IRIS reduction routine suite. This procedure 
relies on derivative estimates and subpixel interpolation to calculate
the locations and amplitudes of the features of the 
\ion{Mg}{2} h \& k lines 
\citep[described in detail in][]{2013ApJ...778..143P}. In this work 
we concentrate our analysis on the properties of the k3 and h3 features,
 which are the central extrema (global maxima or a local minimum) of the line profile,
 that is  
always present, even in the plage region \citep{1967ApJ...149..239T}. The
\ion{Mn}{1} 280.1 nm line is situated between the \ion{Mg}{2} k and h
lines that produce a sloped background continuum. 
We used the IDL routine \emph{gaussian\_fit} to fit a combination of a
Gaussian plus an inclined line on the wavelength range of 
$\pm$ 0.03 nm around the line center because the \ion{Mn}{1} line has a 
regular absorption line shape. We derived the line 
properties from the parameters of the fitted Gaussian profile.
Analysis of the \ion{Mn}{1} 280.1 nm line and \ion{Mg}{2} k feature
formed the basis for the study by \cite{2018MNRAS.479.5512K}, where
the authors found clear signatures of wave propagation 
throughout the quiet solar atmosphere. The 
IRIS spacecraft pointing jitter during the sequences is negligible,
verified by the cross correlation of individual slitjaw frames.

After deriving the fits of the spectral lines and calculating the 
resulting Doppler velocities and line-core intensities,
we cleaned the data from non-converged line fits, which 
amounted to a few percent of the total fits.
We first removed any non-converged fit values 
by replacing them with a 3$\times$3 pixel median filter that
excludes nearby non-converged fits pixels. 
We further smoothed out any discontinuities in the temporal 
domain in the velocity signal which are 
above the local sound speed (7 km~s$^{-1}$) with a 
3$\times$3 pixel median filter, which
corresponds to a 0.5{\arcsec}$\times$ 27 seconds kernel for the 
plage and to a 0.5{\arcsec}$\times$ 51 second kernel for the
internetwork. The Nyquist
frequency of our data is 29 mHz for the internetwork and 
51 mHz for the plage dataset. In the
analysis in Section~\ref{sec:power_spectra} we show that the 
frequencies containing valuable information are between 5 and 20 mHz, 
well below the Nyquist frequency.
The spatial smoothing over 0.5{\arcsec} does not affect the 
estimated wave properties, as 
previous work utilizing high resolution data \citep[e.g.][]{2007A&A...461L...1V}
has shown that the
coherence scale of the velocity signals in the chromosphere 
is of similar spatial scale (see the bottom two rows of 
Figure~\ref{fig:diagnostics_overview}). The resulting data products
from the aforementioned reduction procedures 
are presented in Figure~\ref{fig:diagnostics_overview}.
The left column shows an internetwork region and the 
right one presents a plage region, both
observed near to the disc center. Because
the lower part of the plage field of dataset field of view
is occupied by an internetwork, we exclude this part from 
the plage analysis. In particular, we use the slit locations between 
pixels 400 and 705, which are marked in the right column
of Figure~\ref{fig:diagnostics_overview} as the green (red) lines. 
For the internetwork, we use the full extent of the slit.

\subsection{Processing of the IBIS data}

This study uses data from the Interferometric BIdimensional 
spectrograph \citep[IBIS, ][]{2006SoPh..236..415C} instrument acquired during the 
ALMA coordinated observing campaign on 2017 April 23.
The observed region was centered on the leading edge of AR 12653. The FOV
was 96\arcsec and included regions of 
plage, internetwork, network and penumbra. 
These observations were taken between 17:25-18:12 UT and 
include scans of the \ion{Na}{1} D$_1$ 589.6 nm
and the \ion{Ca}{2}
854.2 nm line, consisting of 24 and 27 points in each line
respectively, which were described in detail in \cite{2022ApJ...933..244H}.
This 
data series has a temporal cadence of 16 sec and 
spectral resolution of at least R $ \gtrsim$ 200,000 
\citep{2008A&A...481..897R}.
The line cores were more 
densely sampled than the wings of the spectral lines 
because the core region is used for deriving the quantities used in this 
study (Doppler velocities and line-core intensities).
The IBIS data processing is described in detail in 
\cite{2019ApJ...881...99M}, where
we have applied the standard reduction techniques of removing instrumental
and atmospheric image aberrations and destretching the resulting data
to the HMI whitelight (atmospheric seeing-free) reference.
In this work we
use the datasets starting at 15:54 UT and 16:37 UT, 
which were taken under conditions of good seeing. 

\section{Properties of the observed power spectra}
\label{sec:power_spectra}

We studied the wave dynamics in the observed chromospheric diagnostics
by analyzing their power spectra. The power spectra are derived for each
pixel in the selected interval along the slit from the squared absolute value of the 
Fourier transform of the time series, giving us the power spectral density (PSD)
of the data. The power spectra 
of the IRIS data exhibit ubiquitous 
power law shapes at frequencies above the acoustic cutoff
present in all chromospheric 
and photospheric observables. These power laws exhibit similar behavior 
to those previously observed in the chromosphere, for 
example in \cite{2008ApJ...683L.207R}  and will be further 
discussed further below.
The average shapes, slopes and other properties of the power laws
are presented in this section.

\label{subsec:obs_power_laws}

\begin{figure}	
		\includegraphics[width=0.45\textwidth]{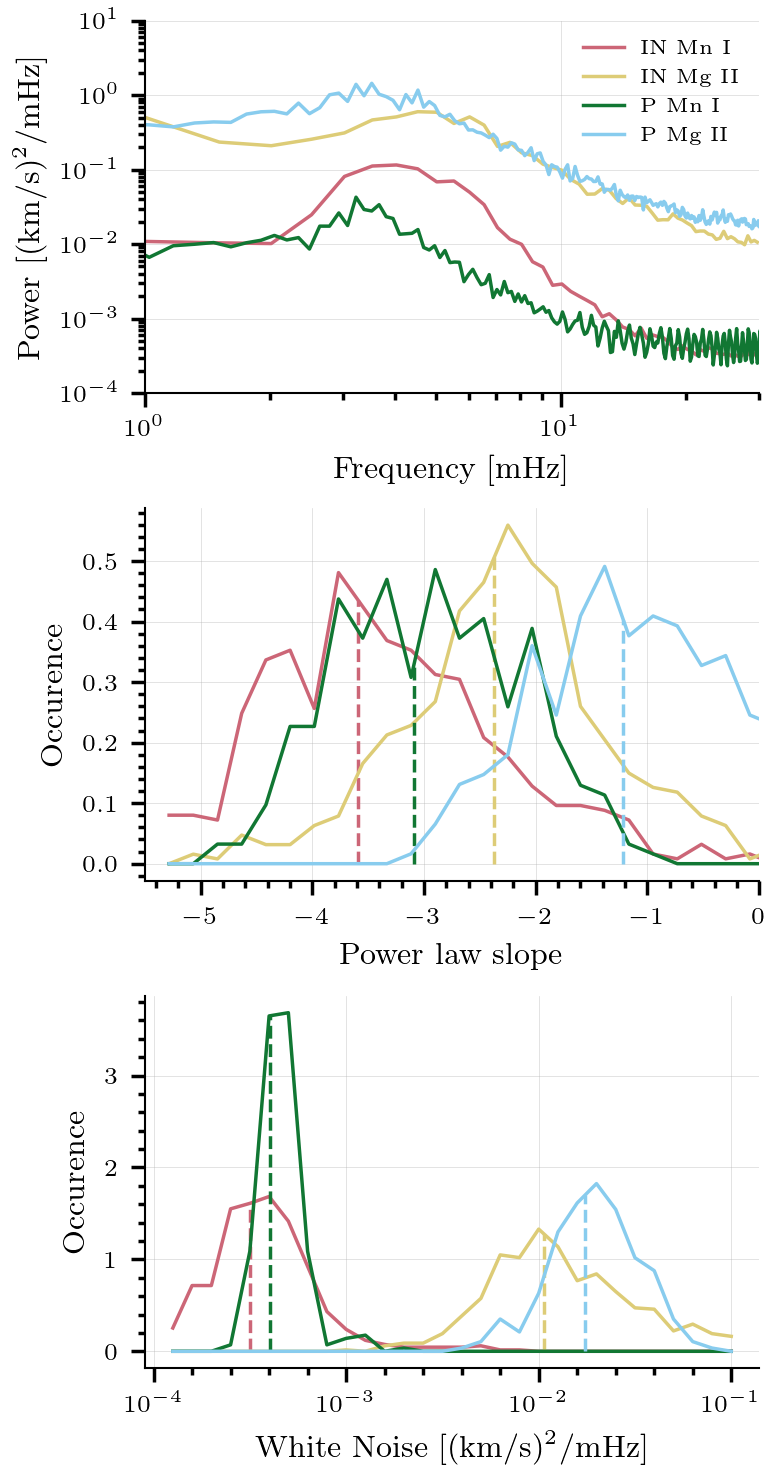}
		\caption{Observed power spectra and their 
			power law properties for the different solar regions and 
			spectral diagnostic. \emph{Top Panel:} Average power spectra.
			{\emph{Middle Panel:}}  Histograms of the slopes of the fitted power laws. 
			{\emph{Bottom Panel:}} Histogram of the white noise floors for the different diagnostics.
			The color coding is consistent throughout the paper.
			The analysis of the data presented in this figure are described in detail in
			Section~\ref{subsec:obs_power_laws}. } 
		\label{fig:PSD_overview}
\end{figure}

\begin{table}[!ht]
	\begin{center}
		
		\label{tab:PSD_props}
		\begin{tabular}{l|c|l|c} 
			\textbf{Solar Feature} & \textbf{Slope} & \textbf{Noise floor} & 
			\textbf{$\langle v^2 \rangle$} \\
			Spectral line & & [(km/s)$^2$/mHz] & [(km/s)$^2$] \\ \hline
			IN \ion{Mn}{1} & $-$3.56$^{+1.02}_{-0.89}$ & 3.2$^{+2.1}_{-1.3}$ 10$^{-4}$ &
			0.18$^{+0.12}_{-0.065}$ \\
			IN \ion{Mg}{2} k3 & $-$2.33$^{+0.85}_{-0.91}$ & 1.1$^{+1.7}_{-0.5}$ 10$^{-2}$ & 
			1.76$^{+0.97}_{-0.55}$ \\
			Plage \ion{Mn}{1} & $-$3.09$^{+0.88}_{-0.81}$ & 4.0$^{+0.9}_{-0.6}$ 10$^{-4}$ &  
			0.02$^{+0.03}_{-0.005}$\\
			Plage \ion{Mg}{2} k3 &$-$1.22$^{+0.90}_{-0.86}$ & 1.7$^{+0.9}_{-0.6}$ 10$^{-2}$  &  
			2.31$^{+1.18}_{-0.98}$ \\					
		\end{tabular}
		\caption{Average PSD
			properties of the observed solar regions in the two IRIS lines
			(\ion{Mn}{1} 280.1 nm and \ion{Mg}{2} k3) with the 10th/90th 
			percentile quoted. The amount of oscillatory velocity power is 
			calculated between 5 and 20 mHz with the white noise subtracted.
			The calculation of the properties is described 
			in detail Section~\ref{subsec:obs_power_laws}.
		}
	\end{center}
\end{table}

Figure \ref{fig:PSD_overview} presents the average 
power spectral profiles (PSDs) and their derived
properties for the different solar regions
and spectral diagnostics.
The average power spectra for the different solar regions 
are shown in the top panel. The internetwork data exhibits the typical
3 minute (5 mHz) peak in both the \ion{Mn}{1} line (lower chromosphere)
and the \ion{Mg}{2} k line (upper chromosphere).
This can be seen clearly from the
last two rows of Figure~\ref{fig:diagnostics_overview}, where the velocity
diagnostics of the quiet sun exhibit regular pattern with the 
time scale of about 3 minutes. 
The plage data exhibits a peak at lower frequencies, around the
3 mHz (5 minute) oscillations
\citep[as previously shown by][]{2004ESASP.547...25D, 2022arXiv220301688M,
	2022MNRAS.512.4164S}, which is more pronounced for the 
lower chromospheric diagnostics.
Furthermore, the Doppler velocity observations in the plage (last two rows of 
Figure~\ref{fig:diagnostics_overview}) do not seem to exhibit 
the clear oscillatory pattern seen in
the internetwork data, which results in a less well defined peak in their Doppler velocity
power spectra.  

\cite{2008ApJ...683L..91Z} have suggested that the
lower frequency peak in the velocity PSD in the plage regions might be 
a signature of the kink wave frequency in the chromosphere. 
However, we did not find a clear correlation between the cotemporal
magnetic field strength in the 
photosphere measured by SDO/HMI \citep{2012SoPh..275..229S}
and the peak of the plage velocity PSD, 
as suggested from the behavior of the kink-wave cutoff. We
intend to extend this study to look for the signatures of the kink-wave
cutoff frequency 
complemented with chromospheric magnetic field measurements 
from DKIST \citep{2020SoPh..295..172R} combined with IRIS observations
in a following publication.

To quantify the usable range of frequencies for our analysis we 
calculated the white noise floor, that is clearly 
seen in Figure~\ref{fig:PSD_overview} Panel (a) as the flat, 
frequency-independent signal at high frequencies. We compute
the white noise floor as the median power
above 25 mHz frequency. 
This noise-frequency cutoff is outside of the frequency 
range used for the wave-power analysis.  
The white noise floor distributions of the different
solar regions are shown in the Figure 
\ref{fig:PSD_overview} Panel (c). Similarly to the results in Paper I,
we observed that the white-noise floor is slightly higher for 
the plage when compared with the internetwork regions. 
We also found that the \ion{Mg}{2}-derived diagnostics have a higher 
white noise floor compared with the \ion{Mn}{1} ones. This trend
might be due to the
measurement technique and/or the nature of the chromospheric lines 
in question, as the \ion{Mg}{2} lines have a complex shape that 
requires an elaborate fitting routine \citep{2013ApJ...778..143P}.
By examining the mean frequency when power rises above the white
noise level, we defined the meaningful frequency region of the PSDs 
to be used for further analysis to be 20 mHz for internetwork
regions and 12 mHz for the plage regions, because
white noise dominates above those frequencies, as clearly
seen in panel (a) of Figure \ref{fig:PSD_overview}.

We perform linear fit on the log-log representation of the 
velocity PSDs to estimate the power law slopes.
The middle panel of Figure \ref{fig:PSD_overview} presents
the power law slopes of the observed PSDs for the different
regions of interest.  For the plage regions we fit
the data between 3 and 12 mHz, and for the quiet Sun we fit the interval
between 5 and 20 mHz, due to the different levels of white noise, 
discussed in the previous paragraph.
The dotted lines show the median of the power law slope
distributions. The power laws of the diagnostics formed in the lower chromosphere
exhibit steeper slopes compared to the ones formed in the upper chromosphere.
Interestingly, for both IRIS lines the plage exhibits steeper power law slopes than the 
internetwork regions, similarly to the behavior of the \ion{Ca}{2} IR line in Paper I. 
The slopes of the vertical velocity PSDs are signatures of the wave environment 
in the chromosphere and they are a crucial dynamic constraint for 
realistic rMHD models of the solar chromosphere.

\begin{figure}

		\includegraphics[width=0.45\textwidth]{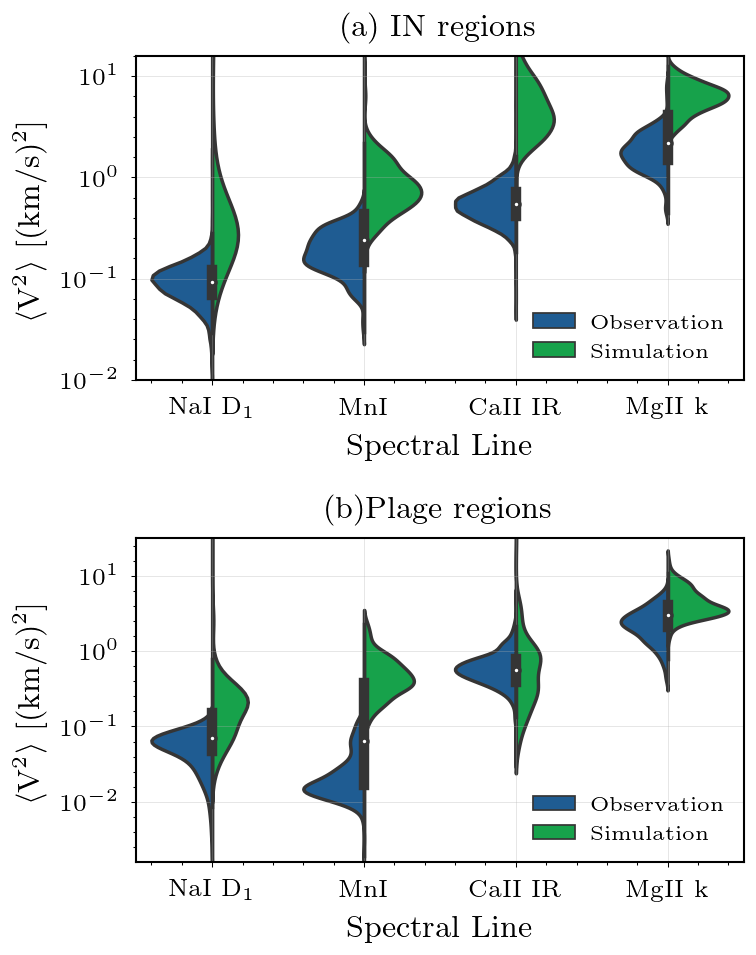}
		\caption{Integrated Doppler velocity oscillation power between 5 and 20 mHz in the 
			IRIS and IBIS diagnostics in the two different solar features
			 (see Section~\ref{sec:power_spectra}), after subtraction of the
			white noise floor. The ordering of the spectral lines 
			reflects their relative average height of formation in the solar atmosphere.
			The blue distributions are real observations, whereas the	 green
			distributions are Bifrost-derived synthetic observables.}
		\label{fig:Vrms_total_power}

\end{figure}

Figure~\ref{fig:Vrms_total_power} shows the 
integrated Doppler velocity oscillatory power between 5 and 20 mHz in the \ion{Mn}{1} 
and the \ion{Mg}{2} k$_3$ features as the blue distributions.
We analyzed only these frequencies,
because waves with these periodicities, above the acoustic cutoff frequency (about
5 mHz in the solar photosphere), will be able to propagate upward. 
The acoustic cutoff frequency varies across the solar atmosphere
\citep{2018A&A...617A..39F,Jefferies_2019}, being lowered at 
locations with strong magnetic field concentrations 
\citep{2011ApJ...743..142H}, but our choice to exclude the power between 3 and 5 mHz
is a conservative estimate of the wave flux, which will not change the end 
result by more than a factor of about two, which is not 
enough to change the conclusions of this work, as shown in Section~\ref{sec:Inferred_flux}. 
We also degraded the resolution of the synthetic data down 
to the resolution of the IRIS and 
IBIS instruments to take into account their diffraction limits.

We also included the amount of oscillatory power from the 
optical lines of \ion{Na}{1} D$_1$ and \ion{Ca}{2} 854.2 nm observed with IBIS.
In all cases, for both IRIS and IBIS diagnostics, we have subtracted a local estimate of 
the high-frequency white noise component for each pixel,
following the noise estimation procedure 
described in the previous paragraph.
Those IBIS observations were obtained on a different day and region than the 
IRIS data analyzed here, but we applied feature-selection criteria described in Paper I,
making for a suitable statistical comparison between these diagnostics.
In Figure~\ref{fig:Vrms_total_power} the blue distributions are 
derived from observations, and the green ones from simulations, which will
be described in Section~\ref{subsec:Bifrost_models}.
The average values of the IRIS velocity fluctuation power 
are summarized in Table~\ref{tab:PSD_props}.

An increase in the amount of velocity oscillatory power is observed
with increasing height in the observations in Figure~\ref{fig:Vrms_total_power}, 
where the spectral diagnostics are arranged in order of increasing height of formation.
This is presumably due to the steeply decreasing density with height in the solar 
atmosphere, leading to increasing wave amplitudes, even 
though the actual wave flux maybe decreasing with height. We also
note that the absolute amount of observed line-of-sight velocity
oscillations is higher in the internetwork in the lower chromosphere compared with
the plage regions. However, in the middle and upper chromosphere the difference in
the amount of velocity oscillation between the internetwork and the plage 
almost diminishes. 

The \ion{Na}{1} D$_1$ line velocity data agree well with the 
velocity data from the \ion{Mn}{1} 280.1 nm line, which is unsurprising
given that both lines are formed at 
similar heights on average \citep{2010ApJ...709.1362L, 2013ApJ...778..143P}. 
The \ion{Ca}{2} 854.2 nm line shows velocity fluctuation amplitudes
 between the \ion{Mn}{1} and \ion{Mg}{2} k3 feature. This confirms that 
the resulting wave amplitudes, coming from different 
spectral lines is self-consistent and presents a uniform physical picture of the 
amount of wave amplitude in the solar atmosphere. 

\section{Acoustic wave propagation in solar simulations: 1D vs 3D models}
\label{sec:Modeling}
The energy flux $F_{ac}$ of propagating acoustic waves with frequencies 
between $\nu_{ac}$ (the acoustic cutoff frequency) and an upper-limit
frequency $\nu_{1}$
can be derived from observations with the following expression 
\citep[following the derivation in ][]{1974soch.book.....B, 2009A&A...508..941B}:

\begin{equation}
F_{ac} = \rho \sum_{\nu' = \nu_{ac}}^{\nu_1} 
\frac{\left \langle v_{obs}^2(\nu') \right \rangle}{{\cal{T}}^2(\nu')} v_{gr}(\nu') 
\label{eqn:wave_flux}
\end{equation}
where $\rho$ is the plasma density at the formation height of the
observed diagnostic; $\left \langle v_{obs}^2(\nu') \right \rangle$ is the 
observed velocity variance at frequency bin $\nu'$; $\cal{T(\nu')}$ is 
the attenuation coefficient due to the finite thickness of 
the formation region of the spectral line \citep{1980A&A....84...96M}; 
$v_{gr}(\nu')$ is the group velocity of the wave mode at frequency $\nu'$. 
To estimate the wave energy flux,
we have to evaluate the terms on the right side of 
Equation~\ref{eqn:wave_flux} from models or observations. The quantity
$\left \langle v_{obs}^2(\nu') \right \rangle$ can be obtained from the observations as
described in Section~\ref{sec:Observations}. The other 
three quantities however need to be estimated
from numerical models, as we describe in this Section.

We extend previous analyses 
\citep{2005Natur.435..919F,2002A&A...395L..51W,2016ApJ...826...49S}
to compare the differences between 1D models 
(FAL, RADYN) and 3D models (Bifrost). 
In comparison with 1D models, the Bifrost model includes additional physical
processes (dynamical evolution, shock formation, detailed radiative transfer
and non-equilibrium ionization of hydrogen) which produce a
wealth of small-scale phenomena. This approach allows 
for self-consistent description of 
wave propagation in the chromosphere, avoiding some of the problems with 1D 
modeling described in \cite{2005ApJ...631L.155U}.
Previous work by \cite{2021RSPTA.37900170F} compared the 
general wave propagation properties in 3D MHD simulations, including Bifrost,
and found a lack of general agreement among the different models. We 
note that those authors did not explore the observational 
signatures of high-frequency wave propagation
in the chromosphere, which is the central topic of this paper.

\subsection{RADYN models}
\label{subsec:RADYN_models}

We use the same RADYN \citep{1992ApJ...397L..59C, 
	2005ApJ...630..573A, 2015ApJ...809..104A}
runs presented in Paper I to interpret the IRIS observations.
The initial RADYN atmospheric model 
used was an IN atmosphere model with 191 grid points. The model has
a piston-like lower boundary condition
that acts as a sub-photospheric wave driver and an 
open upper boundary with constant temperature of 1~MK. 
RADYN self-consistently solves the equations of radiative transfer, statistical
equilibrium, and the hydrodynamic equations, where the code can take into 
account the time dependent ionization. Furthermore, the RADYN code 
treats in non-LTE the transitions of hydrogen, calcium, and helium with 
6\-, 6\-, and 9\- level atom models respectively.

We synthesized time series of spectral line profiles from these models. 
Based on these synthetic observables, we estimated the line
displacements and intensities described in the further analysis.
To synthesize the \ion{Mn}{1}, and the \ion{Mg}{2} spectral diagnostics 
studied throughout this work, we 
use the RH 1.5D code \citep{2001ApJ...557..389U, RH_15D}. For the synthesis of the 
\ion{Mn}{1} 280.1 nm line, we used the Kurucz line list
database\footnote{http://kurucz.harvard.edu/linelists.html} 
\citep{2018ASPC..515...47K} and we synthesized
it in local thermodynamic equilibrium (LTE). We
note that the lines of \ion{Mn}{1}
exhibit non-LTE effects \citep{2019A&A...631A..80B}, but 
we leave the assessment of importance of these effects for a future work.
To synthesize the \ion{Mg}{2} h \& k lines we used the RH code 
in non-LTE mode with 10 plus one \ion{Mg}{3} ground levels and PRD treatment 
\citep[the same setup used in][]{2012A&A...543A.109L}.

\subsection{Bifrost models}
\label{subsec:Bifrost_models}

\begin{figure*}
	\includegraphics[width=\textwidth]{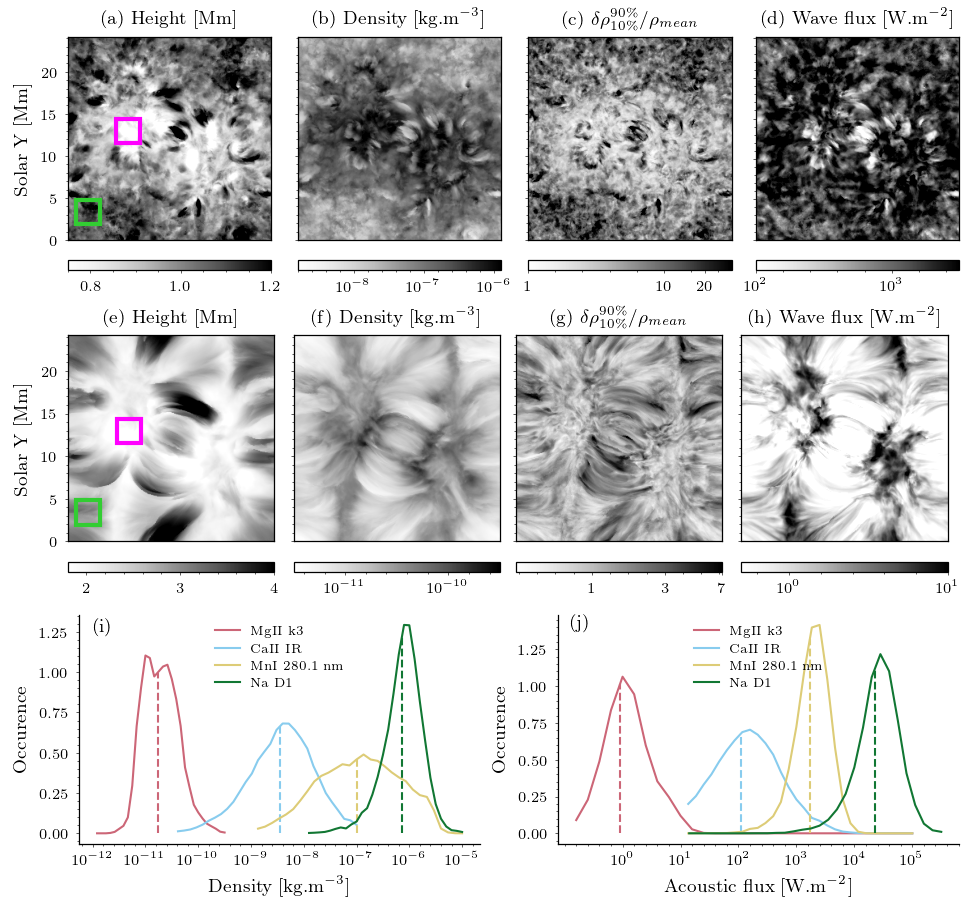}
	\caption{Results 
		from the Bifrost spectral synthesis. The top row 
		shows diagnostics derived
		from the \ion{Mn}{1} 280.1 nm line and the middle row 
		shows those for the \ion{Mg}{2} k3 feature. Panels (a) 
		and (e) show the time-averaged height of optical
		depth unity of the line core, panels (b) and (f) the
        time-averaged density at optical depth unity for the line core;
		panels (c) and (g) show the ratio of the 
        plasma density change over time to the mean plasma density 
        at the formation height of the spectral lines, and 
		panels (d) and (h) the acoustic flux at the
		formation height of the spectral line.
		The green and magenta 
		squares in panels (a) and (e) are the representative regions
		that we equate to internetwork and plage regions in our observables 
		in the following analysis.
		The bottom left panel (i) shows the distributions of the density
		at the height of formation in
		the simulation for the different diagnostics;
		the bottom right panel (j) shows the distributions of the 
		acoustic flux at the height of formation of the diagnostics. } 
	\label{fig:Bifrost_overview}
\end{figure*}

Modern 3D radiative magnetohydrodynamic (rMHD) codes appear to 
result in increasing levels of realism of the simulated 
solar atmosphere \citep{2019A&A...631A..33B}. To leverage the
advantages of multidimensional rMHD simulations,
we use the publicly available Bifrost
\textbf{data cubes}\footnote{Available at \url{http://sdc.uio.no/search/simulations}.} 
of enhanced network \emph{en024048\_hion} 
\citep{2011A&A...531A.154G, 2016A&A...585A...4C}. In this simulation we 
utilize the quiet regions as realizations of quiet Sun internetwork and
the more active network as representative of plage regions.
We further employed the publicly available radiative transfer products for the 
\ion{Mg}{2} h \& k and the \ion{Mn}{1} 280.9 nm lines which are synthesized
with the RH 1.5D code and publicly available for the enhanced network
\emph{en024048\_hion} \citep{2013ApJ...778..143P}.
We also synthesized the \ion{Ca}{2} 854.2 nm and
\ion{Na}{1} D$_1$ lines with RH 1.5D in nLTE. 
We used a 6-level model atom for the \ion{Ca}{2} 854.2 nm line,
including a \ion{Ca}{3} ground state; for the \ion{Na}{1} D$_1$ line,
we used a model atom with 4 levels which includes a \ion{Na}{2} ground state. 

Before proceeding with the analysis of the spectral synthesis products, 
we note a few deficiencies of the Bifrost models, which should be 
kept in mind while interpreting the following results.
First, the UV solar spectrum is not well reproduced, with spectral features lacking in 
intensity and width \citep{2013ApJ...778..143P}. 
As discussed previously in \cite{2016A&A...585A...4C}, 
this might be due to a combination of factors, like insufficient 
heating in the chromosphere and corona and the lack of small-scale 
motions in the simulated atmospheres.
The other major drawback of these models is the presence of 
global oscillations over the whole simulation domain 
with velocity perturbations on the order of a few km/s
in the lower chromosphere, accompanied by density fluctuations on the order of
20 percent
\citep[described previously in][]{2016A&A...585A...4C, 2021RSPTA.37900170F}. 
We have attempted to remove the signature of these
wave modes in our analysis by filtering them in temporal Fourier space, 
given their periods are lower (about ten minutes) compared with
the periods of interest in this paper and are coherent over the whole domain.

\subsection{Properties of the synthetic observables from Bifrost} 
\label{subsec:comparison_models} 

Figure~\ref{fig:Bifrost_overview} shows the formation properties
of the \ion{Mn}{1} 280.1 nm line (top panel) 
and \ion{Mg}{2} k3 (middle panel) in the enhanced network Bifrost simulation.
Panels (a) and (e) show the height of optical depth unity, referred to as 
the \emph{height of formation} of the spectral line. These panels indicate that the  
two spectral lines are 
formed at significantly varying heights in the atmosphere at 
different locations in the FOV, 
as previously shown in \cite{2013ApJ...778..143P}. This spread of the height 
of formation is a significant contribution to the broad distribution of 
densities at the $\tau$=1 heights,
which are shown in Panels (b) and (f). This raises the
question of the applicability of the approach based on inferring 
the acoustic flux using a singular density value for a given spectral line.  
Additionally, the effective height of formation of a 
spectral line may change as the atmospheric properties evolve in time.
The amplitude of this effect is illustrated in 
panels (c) and (g) which show for each line the ratio of the  
difference between the 10th and 90th percentile of the temporal density variation 
to the time-averaged plasma density at the height of formation 
for each pixel. We note a strong temporal variation of the 
density at the height of formation with time on the order of a factor of few for the
same temporal location, similarly to 
\cite{2023arXiv230103273F}. This change
is due to the passing wave fronts and the different amplitudes 
are due to the significantly different properties of formation of the diagnostics
in the two regions.  For the \ion{Mn}{1} line, the density of 
formation changes by an order of magnitude in internetwork regions over time, 
but relatively less in the enhanced network regions.  
For the \ion{Mg}{2} k line, we see that the density
changes most significantly 
along the fibrilar structures, connecting the two magnetic regions in 
the simulation domain.

Based on the spectral synthesis of the two UV and the two optical lines, 
we computed $\tau$=1 plasma density histograms for each spectral line 
from the first snapshot of the simulation.
Panel (i) shows that the densities at the $\tau$=1 heights exhibit 
wide distributions that present a challenge 
for the computation of the wave fluxes. 
If we examine indicative enhanced network (magenta squares
in Panels (a) and (e)) 
and internetwork (green squares in Panels (a) and (e)) structures
we find that those regions exhibit almost constant density inside the small boxes. 
The average of the density from those regions 
could be used as  the representative of the values to be used in 
Equation~\ref{eqn:wave_flux} when estimating the acoustic fluxes. This 
strong dependency of the 
density on the particular solar region, further described in Section~\ref{sec:Uncertainties}, makes providing 
an accurate model for every solar feature crucial for the accurate 
estimation of the wave flux. 

The acoustic wave flux present in the simulation cube can
 be computed at different heights
as the plasma conditions are known. Due to the varying formation conditions 
of the diagnostics, described in the previous paragraph,
we estimated the average height of formation for each spectral line
separately for each column of the simulation. Based on the average height 
of the column, we extracted the average plasma density and 
the amount of vertical velocity oscillatory power 
between 5 and 20 mHz at that height in the simulation.
Based on these estimates, we computed the average wave flux 
at the local formation height of the spectral lines.
The resulting acoustic wave-flux distributions for all spectral lines
are presented in Panel (j) of 
Figure~\ref{fig:Bifrost_overview}. The amount of acoustic flux with height
decreases significantly, in contrast to the almost constant amount of wave flux
in the 1D RADYN chromosphere \citep{2006ApJ...646..579F}.
This is the typically observed behavior of the wave flux with height, 
as hinted by previous observations \citep[e.g.,][]{2020A&A...642A..52A}.
The amount of acoustic wave flux in the
Bifrost simulation chromosphere resembles the results based on the RADYN models
in Paper I, but exhibit a more realistic decrease of the wave flux with height
\citep{2005ApJ...631L.155U}.
This is further described in detail in Section~\ref{subsec:comparison_models}.

Based on the spectral profiles computed from the Bifrost simulation, we measured
the Doppler velocities using the same procedure as for 
the real observations, described in Section~\ref{subsec:IRIS_data_processing}.
We have also subtracted a white-noise estimate, 
derived as frequency independent  
at high frequencies of the power spectrum.
We compare the amounts of Doppler velocity fluctuations in the real 
data (blue distributions) and the simulations (green distributions)
in Figure~\ref{fig:Vrms_total_power}. The data for the simulation results are 
based on the aforementioned green and pink regions in 
Figure~\ref{fig:Bifrost_overview}. The simulations seem to 
exhibit significantly higher velocity oscillation power 
than the actual Sun, often up to a magnitude more.

\begin{figure}
		\includegraphics[width=0.475\textwidth]{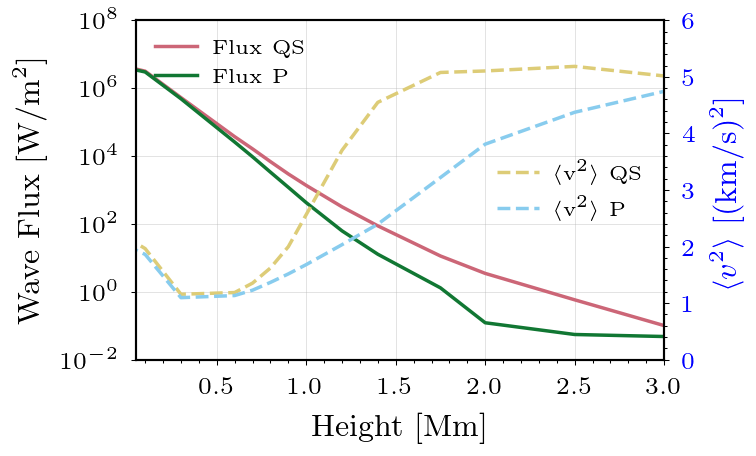}
		\caption{Height variation of the
			 acoustic flux and vertical velocity fluctuation amplitude between 5 and 
			20 mHz in the Bifrost model for internetwork (green) and plage regions (red). 
			The solid (dotted) lines show the wave flux (velocity fluctuation). The regions 
			of the simulation used are shown in Panels (a) and (e) of
			Figure~\ref{fig:Bifrost_overview} as the colored squares.}
		\label{fig:Bifrost_v_rms}
\end{figure}

The acoustic wave energy propagating through 
the chromosphere appears to be mostly dissipated 
by the time it reaches the height of formation 
of the \ion{Mg}{2} k3 feature (see Figure~\ref{fig:Bifrost_overview} Panel (h)).
in the simulations. This is further illustrated in 
Figure~\ref{fig:Bifrost_v_rms}, where the height dependence of the 
acoustic flux in the internetwork and magnetic 
concentration regions are shown.
The amount of velocity variance is also shown in 
Figure~\ref{fig:Bifrost_v_rms} as the dashed lines. We can see that the 
amounts of vertical velocity oscillatory power in the internetwork and the 
plage are similar in the photosphere,
but in the chromosphere the internetwork has higher velocity oscillation power
by a factor of two. 
However, when taking into account the slightly 
lower density at chromospheric heights 
for the internetwork, compared with the enhanced network, we found
that the velocity amplitudes are almost the same over a large range of heights. 

This analysis shows the drawbacks of using 1D atmospheric models
to infer the wave fluxes. First and foremost, perturbative approaches 
\citep[such as][]{2009A&A...508..941B,2020A&A...642A..52A} cannot account for 
the atmospheric properties changing significantly between different
solar features. There has been previous work by 
\cite{2005Natur.435..919F,2006ApJ...646..579F} that used
time-dependent 1D HD RADYN models to infer wave fluxes from the
TRACE observations, but these authors did not use 
differing starting atmospheric models to study the behavior of 
different solar features; or have used multiple 1D semi-empirical static
models \citep{2016ApJ...826...49S}.
Furthermore, the analysis of the 3D models shows that the 
high-frequency waves do not over-saturate the chromosphere with 
acoustic power as in the 1D case \citep{2005ApJ...631L.155U}.

A key property that affects the estimation of the acoustic
flux is the plasma density.
We do not argue
about the veracity of the conclusions in either approach, as the reliability
of 3D models to represent the wave dynamics of the 
solar atmosphere is still under debate
\citep{2021RSPTA.37900170F}. 
Furthermore, the too-weak spectral lines in the synthetic spectra are most probably due to 
low densities in the Bifrost simulations \citep{2016A&A...585A...4C}.
In the next section, we compare the different modeling approaches, quantifying their systematic differences, which might explain some of the discrepancies among the previous results for acoustic flux estimates.

\section{Systematics of acoustic wave flux estimation from 3D vs 1D models}
\label{sec:Uncertainties}

\begin{figure*}
	\includegraphics[width=\textwidth]{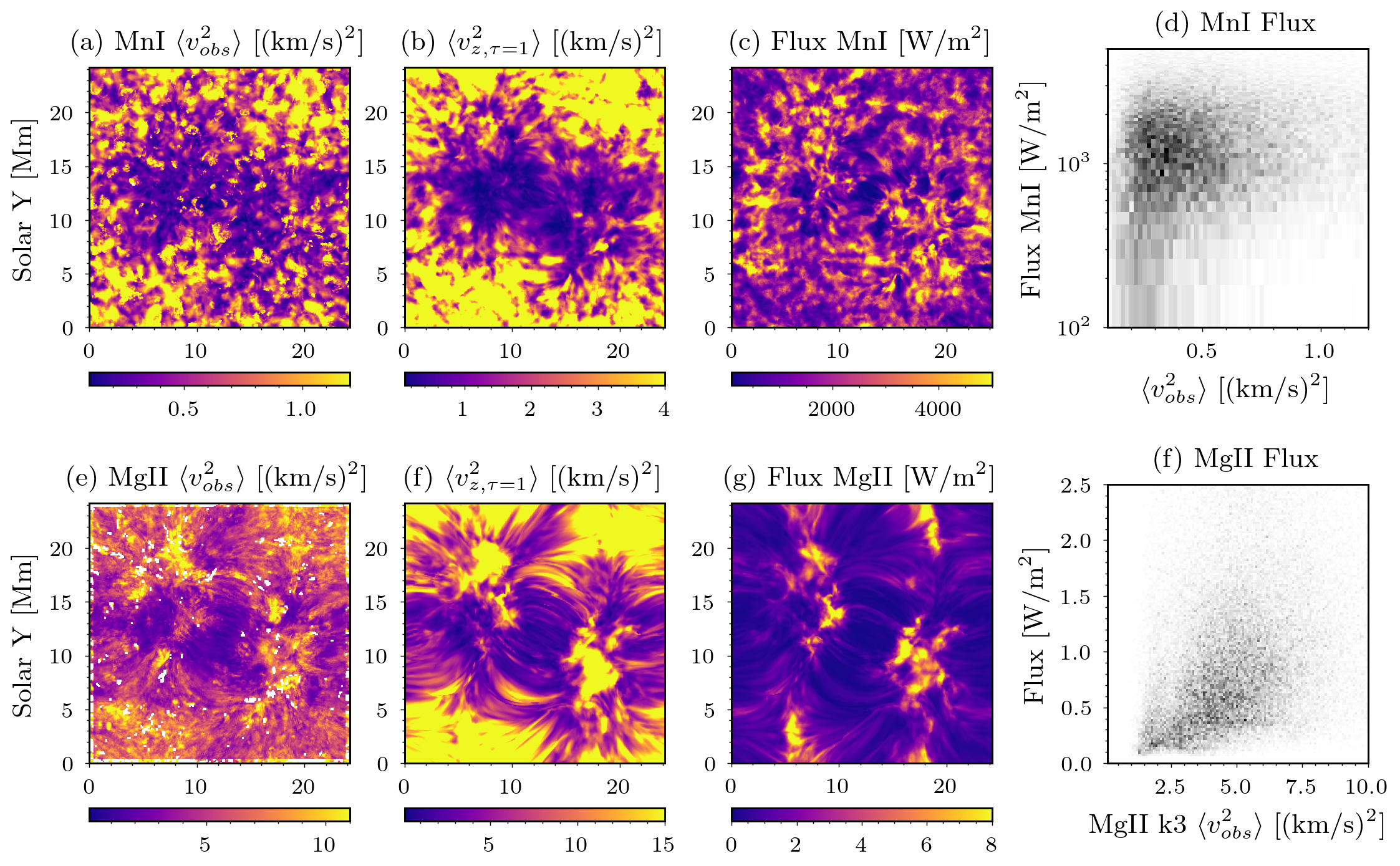}
	\caption{Comparison of the
		synthetic Doppler velocity fluctuations in the \ion{Mn}{1} line and
		\ion{Mg}{2} k3 feature and the wave fluxes at the corresponding
		heights in the model atmospheres.
		The top presents the following \ion{Mn}{1} 280.1 nm derived diagnostics:
		panel (a) shows the measured synthetic Doppler 
		velocity fluctuations between 5 and 25 mHz; 
		panel (b) shows the vertical velocity oscillatory power between 5 and 25 mHz in the 
		Bifrost simulation at the $\tau=1$ height for each column;
		panel (c) shows the acoustic flux as measured 
		in the simulation at the $\tau=1$ height for each column, and 
		panel (d) shows the scatter plot between the quantities in
		(a) and (c). 
		The bottom row shows the same features, but for the \ion{Mg}{2} k3.}
	\label{fig:Bifrost_v_rms_vs_flux}
\end{figure*}

The spectral synthesis of observables from numerical solar 
models provides us with a 
direct way to examine how the variations in measured diagnostics relate to the 
actual changes in atmospheric plasma properties.
In this section we examine the behavior of the 
following components of Equation~\ref{eqn:wave_flux}
in different modeling approaches: \emph{(i)} atmospheric velocity at 
height corresponding to the observed Doppler velocity 
measurement and what is the source of the observed velocity
fluctuations -- true plasma motions or rather
changes in the $\tau$=1 surface;
\emph{(ii)} what is the density at the height of formation 
associated with the oscillatory signal;
and \emph{(iii)} what is the transmission coefficient in different regions of the solar atmosphere.
We compare the results from the 3D Bifrost simulations with 
previous results from RADYN and FAL atmosphere-based
modeling \citep{2011JGRD..11620108F}. 
Such comparison allows for estimating the systematic errors that are 
introduced by using a particular modeling approach. This is an important aspect 
of these studies that has not been well constrained previously. 
We demonstrate that specific choices made for the height of formation, 
density, and transmission coefficient can drastically change the 
conclusions from these of studies.

\subsection{Measuring velocity fluctuations, but where?}
\label{subsec:uncertainty_vel_fluc}

The analysis in Section~\ref{subsec:comparison_models} shows that the
Doppler velocity signals derived from synthetic spectral lines originate from a 
height that can change with time and depending on the underlying solar
feature. Hence, we need to determine at which height the Doppler velocity 
samples the true vertical velocity field most closely.

To constrain to which height the observed Doppler velocity relates to, 
we calculated the Pearson correlation coefficient
between the observed Doppler
velocity and the plasma vertical velocity. 
The highest correlation coefficient values were found at the heights of
the time-averaged optical depth unity which confirmed our previous calculations.

We compared the Doppler velocities in 
the synthetic spectral observations with the acoustic flux at the 
$\tau=1$ height of formation of the spectral line. The results are shown
in Figure~\ref{fig:Bifrost_v_rms_vs_flux}, where the first row is for the \ion{Mn}{1}
line and the second row is the \ion{Mg}{2} k3 feature. 
Optimally, there would be a 
direct mapping between $v_{obs}^2$ and the wave energy flux, which would 
imply that the estimation of the density and the attenuation coefficient 
should be straightforward.

For both spectral lines, there is a good agreement between the 
distribution of the observed synthetic velocity oscillations and the true 
vertical velocity oscillations at the line height of 
formation in the solar atmosphere. Panels (a) and (b) of 
Figure~\ref{fig:Bifrost_v_rms_vs_flux}
show the observed Doppler velocity and the velocity at the time-averaged
$\tau$=1 height for the \ion{Mn}{1} 280.1 nm line. 
Panels (e) and (f) show the same for the 
\ion{Mg}{2} k3 feature. On average the observed Doppler velocity fluctuations are lower
than the true vertical plasma velocities in the solar atmosphere.
This is due to a Doppler velocity attenuation effect that
smears out the vertical velocity signal in the solar atmosphere. 
It is caused by a combination of
multiple phases of the acoustic waves might be present in 
the width of the formation region as well as the
changing line height of formation \citep{1980A&A....84...96M}. This observed 
decrease of the wave amplitudes is described by 
the $\cal{T}$ coefficient, discussed 
further in Section~\ref{subsec:T_coeffs}.
The total amplitudes of the velocities
derived from the synthetic 
observables are on average lower by a factor of two to four,
due to the attenuation 
of the signal.

However, when we compute the acoustic fluxes at the time-averaged
$\tau=1$ surfaces 
of the simulations, we see that the correspondence with the velocity 
amplitudes is mostly nonexistent 
(panels (c) and (g)). This is due to the fact 
that the other major component of the acoustic flux calculation is the density.
The density at the height of line formation
varies significantly in the different regions of the chromosphere, as shown in 
panels (b) and (f) in Figure~\ref{fig:Bifrost_overview}. In particular, these
subfigures show us that the local density changes by 
more than an order of magnitude between
the quiet and enhanced network regions. This can be understood as 
in the hotter (network) regions, the diagnostics are formed at a lower height and 
on average at higher column mass \citep{2011JGRD..11620108F}.
The density variation is significantly higher than
the variation of the amplitudes of the observed velocity fluctuations in the
simulations.

This strong spatial variation of plasma properties results in the poor correlation
between the observed synthetic velocity oscillation power
and the acoustic flux at the line formation region, as shown in panels 
(d) and (f) of Figure~\ref{fig:Bifrost_v_rms_vs_flux}.
The correlation is marginally better for the case of the 
\ion{Mg}{2} k3 feature.  The relatively smaller
change of the density of formation in the case of the upper chromospheric
\ion{Mg}{2} k3 leads to a better correlation 
between the synthetic observed velocity fluctuations and the 
acoustic flux in the atmosphere. 

The conclusion from Figure~\ref{fig:Bifrost_v_rms_vs_flux} is that the 
variations in the
formation height of the spectral lines in different 
features is a significant effect when estimating
the wave flux in the solar atmosphere,
as that will alter the observed wave velocity amplitudes and local densities.
Using fixed values for the density 
will produce results that do not correspond to the true flux at the formation 
region of the spectral lines. Optimally, we would take this into account when
estimating the acoustic flux by employing different densities.
However, as we discuss in the following section, sufficient knowledge of the
local densities in the chromosphere and their range of variation, is still lacking.

\subsection{Chromospheric density estimates are model dependent}

\begin{figure}[htp!]
	\begin{center}
		\includegraphics[width=0.45\textwidth]{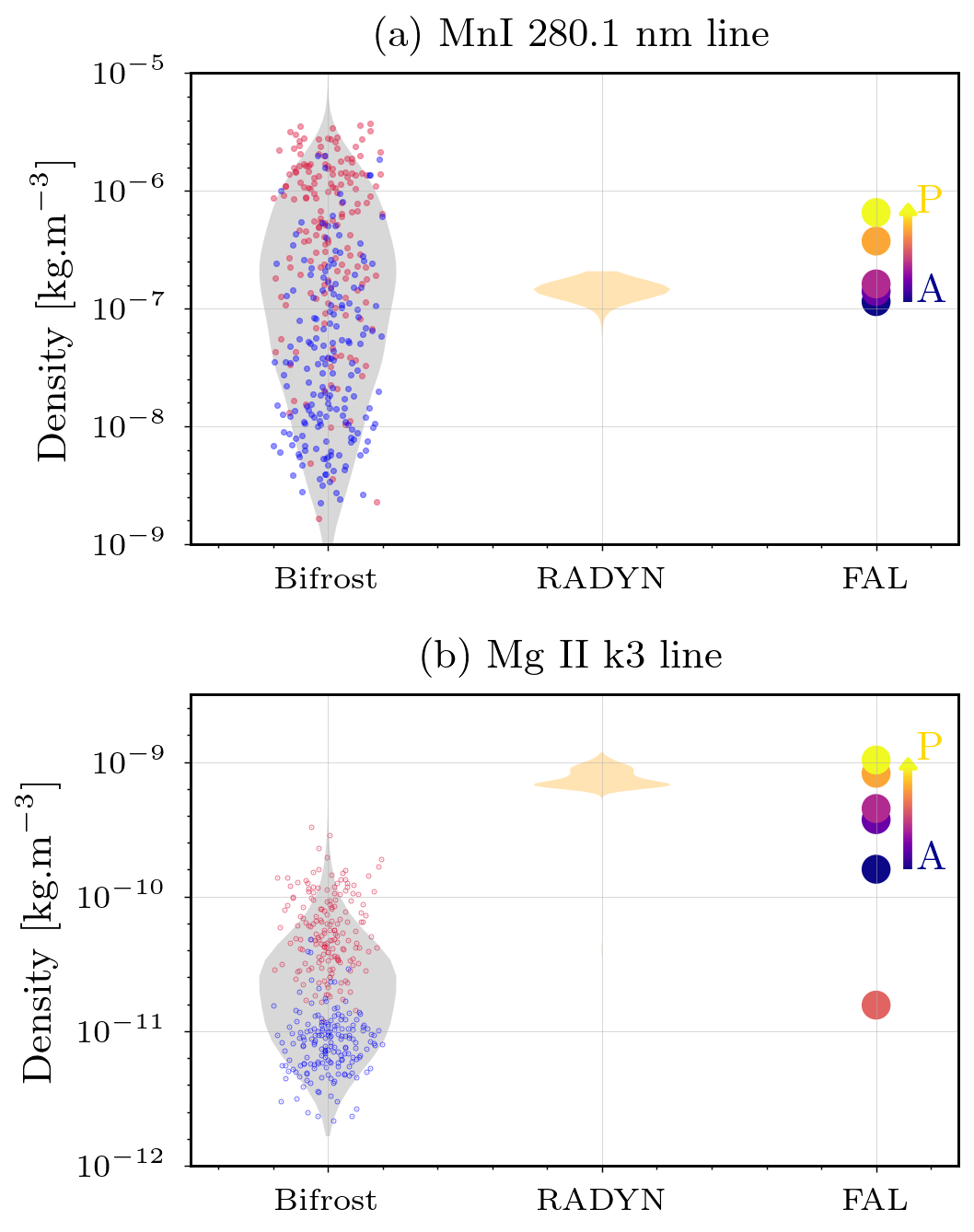}
		\caption{Density at the line formation height for the
			\ion{Mn}{1} 280.1 nm line and the \ion{Mg}{2} k3 
			feature from different wave modeling approaches labeled on the
			abscissa. The top panel (a)
			shows the results for the \ion{Mn}{1} line and the bottom panel (b)
			for the \ion{Mg}{2} k3 feature. The data points overlaying
			the Bifrost density distribution correspond to internetwork (blue) and 
			enhanced network (red)
			regions shown in Figure~\ref{fig:Bifrost_overview} 
			sampled every 200 seconds.
		} 
		\label{fig:rho_comparison}
	\end{center}
\end{figure}

Density is the quantity with the  
highest degree of variability in estimating the acoustic flux in the chromosphere, 
due to its highly corrugated and dynamic structure
\citep{2019ARA&A..57..189C}. 
As alluded in the previous subsection, the density at the
formation location 
of the same diagnostics in different regions of the solar atmosphere
changes by a few orders of magnitude, as illustrated in panels (b) and (f) in 
Figure~\ref{fig:Bifrost_overview}.
In this section we discuss the intrinsic variability of the plasma density 
at the height of spectral line formation in the different modeling 
approaches. The variability described here is due to the different line formation 
conditions in the model atmospheres, not the intrinsic changes due to 
the wave perturbations per se.

Figure~\ref{fig:rho_comparison} shows the distribution of
 plasma densities at the 
$\tau=1$ surface for the \ion{Mn}{1} 280.1 nm line in panel (a) and 
the \ion{Mg}{2} k3 feature in panel (b) for different modeling approaches. 
The three different models 
described here are: Bifrost 3D rMHD simulations described in 
Section~\ref{subsec:Bifrost_models}; the RADYN models described
in Section~\ref{subsec:RADYN_models}; and the FAL11 semi-empirical
1D hydrostatic models, described in \cite{2011JGRD..11620108F}. 
We use the latest FAL models, since they reproduce the average
solar spectra to the best extent, but are in essence very similar to other 1D 
semi-empirical atmospheric models used in previous acoustic wave 
studies.

For the Bifrost rMHD model, we extracted the corresponding densities at
every 5th spatial pixel in both spatial dimensions 
at 200 second intervals. 
The distribution of the Bifrost densities are presented as the gray 
distribution in Figure~\ref{fig:rho_comparison}. We also calculated 
the densities at the two regions of internetwork and active network, marked as the
squares in Figure~\ref{fig:Bifrost_overview}. We plotted them over the full
Bifrost distribution with the blue (internetwork) and red (active network) markers. The 
formation of the lines in the active network is at higher average plasma densities,
which agrees with the previous discussion in Section~\ref{sec:Modeling}. 
For the RADYN models, we calculated the \ion{Mn}{1} and the \ion{Mg}{2} 
lines for the \textit{model\_3000} run from Paper I for every temporal
step, where we have excluded from the synthesis the relaxation
time of the simulation. We calculated the density from the other 
models presented in Paper I that have increasing wave
strength, but the results were
similar to the ones presented here. The FAL models A-P, increasing in 
activity from very quiet internetwork to plage core, are shown with the
colored circles on the right. The relative warmth of the color of the marker 
signifies increasing activity level.

The different modeling approaches produce very different
estimates for the plasma density at the line formation region, as shown in 
Figure~\ref{fig:rho_comparison}. In particular, the Bifrost models
exhibit a high level of intrinsic variation of the density in the different solar features.
 In the case of the \ion{Mn}{1} line, the RADYN-derived density corresponds to 
the quietest FAL models, which is not surprising, given the initial
RADYN atmosphere was based on a relaxed FAL B like model.
Comparing the Bifrost density estimates with the 1D model ones,
we observed that mostly the active network regions 
have mostly similar density to the ones retrieved from the FAL-based modeling. 
In the internetwork, the Bifrost models estimated that the
density of formation is significantly lower than the one derived from the FAL
models, but at some points they exhibit high densities, similar to the ones seen in the
enhanced network.

In the case of the \ion{Mg}{2} lines, the RADYN models exhibit densities
closer to those of the hotter FAL models, opposite from what is 
seen in the \ion{Mn}{1} 280.1 nm line case. However, the more self-consistent
Bifrost simulation exhibits significantly lower density than either
FAL and RADYN simulations for both enhanced network and internetwork. 

Using the values of Bifrost simulation-derived densities for  
flux estimates would lead to lower inferred
chromospheric wave fluxes when compared with using 
values based on the FAL models. We do not dispute which density 
values are more accurate, as the Bifrost models still lack 
heating and sufficient density
in the chromosphere to properly reproduce the observed 
spectral profiles. Instead, we 
highlighted the systematic biases in different wave flux estimates
based on the models used. The large spatial and temporal spread in 
densities in the more dynamic rMHD models does further indicate the use of a 
single or few values of density in computing wave fluxes is likely an
oversimplification that leads to significant uncertainties.

\subsection{Uncertainty of the transmission coefficient}
\label{subsec:T_coeffs}

\begin{figure*}[!htbp]
	\begin{center}
		\includegraphics[width=\textwidth]{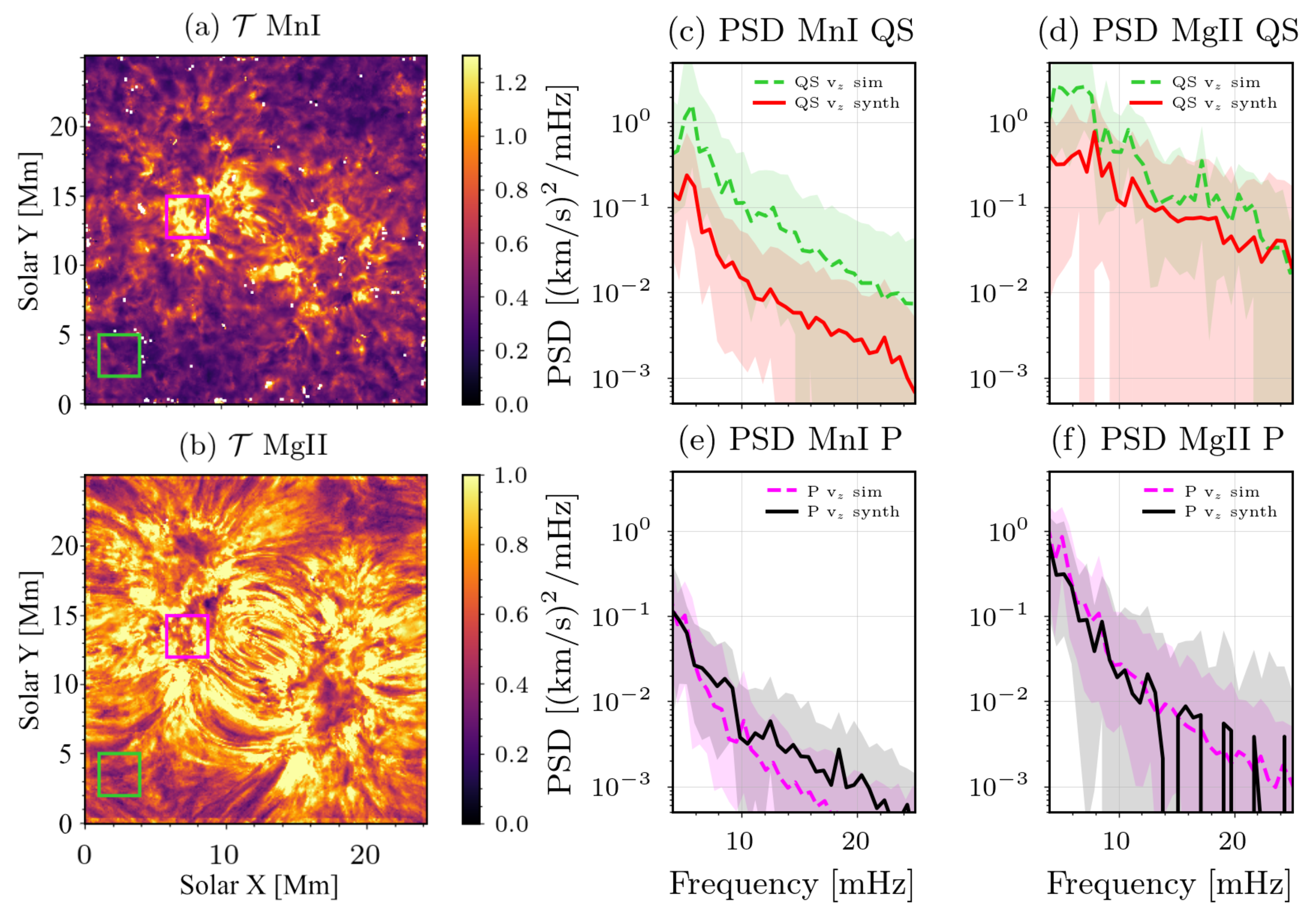}
		\caption{Attenuation coefficient $\cal{T}$ maps in the Bifrost simulation for vertical
			velocity fluctuations between 5 and 20 mHz (panels (a)-(b))
            based on the power spectral density of the observed and true 
            plasma diagnostics (panels (c)-(f)).
			The top panel (a) shows the attenuation coefficient in the \ion{Mn}{1} 280.1 nm line
			and the bottom panel (b) shows it for the \ion{Mg}{2} k3 feature. The green and magenta regions correspond to 
            the dark internetwork and 
			enhanced network regions of interest. Panels (c)-(f)
            compare the power spectral
            density of the vertical velocity oscillatory power in the atmosphere 
            at the formation height of the diagnostic (v$_{sim}$) and the Doppler velocity
            oscillatory power in the synthetic observations (v$_{synth}$) for the quiet Sun 
            and plage regions. Note that the 
            solid/dashed lines are the average of the distributions outlined as the colored
            squares in panels (a) and (b) and the shaded regions are the 10\% / 90 \% 
            percentiles of the distributions. 
		}
		\label{fig:T_coeff_Bifrost}
	\end{center}
\end{figure*} 

The attenuation coefficient $\cal{T}$ is the last model-dependent parameter
in estimating the wave flux. We define $\cal{T}$ in this work as the ratio of the 
standard deviation of the observed and actual atmospheric vertical velocities. We 
take a frequency-averaged approach, as previous work in Paper I
calculated $\cal{T}$ as a function of frequency and 
showed that most of the power is at the lower frequencies.
To examine its variation in the Bifrost simulations,
we calculated the ratio of the standard deviations of the  
Doppler velocities, derived from the synthetic observations, and the 
vertical velocities in the simulation at the time-averaged height of the 
$\tau = 1$ surface. We have filtered the vertical velocities in Fourier space leaving the
frequencies between 5 and 20 mHz.
We adopt an averaging of the velocity fluctuation power 
over the frequency domain for calculating
$\cal{T}$, different from previous work. This makes it more resistant to
noise at the high-frequency limit, which can contribute to the observed Doppler 
velocities solely due to measurement errors. 
The attenuation coefficient maps for both \ion{Mn}{1} 280.1 nm (panel (a))
and \ion{Mg}{2} k3 (panel (b)) are 
presented in Figure~\ref{fig:T_coeff_Bifrost}.
The attenuation coefficient varies significantly over the simulation domain 
and is partially correlated with the type of underlying solar features. 

For the \ion{Mn}{1} 280.1 nm line the attenuation coefficient 
is on the order of $\sim$~0.4 in the internetwork, which might be expected due to 
the strong variation in the of heights being
sampled of the Doppler velocity, as discussed 
in Section~\ref{subsec:uncertainty_vel_fluc}. In the case of the network regions,
the attenuation coefficient is closer to unity due to the fact 
that in these regions the height of formation changes significantly less, as shown in 
Figure~\ref{fig:Bifrost_overview} panel (c).

For the \ion{Mg}{2} k3, the attenuation coefficient is in general
higher compared with the \ion{Mn}{1} lines. In the quieter 
regions the attenuation coefficient is lower (about $\sim$ 0.6) and closer to unity 
in the network regions. In the case of the \ion{Mg}{2} k3 (panel (b) 
of Figure~\ref{fig:T_coeff_Bifrost}),
we see that the extended fibrilar structures in the simulation
are clearly correlated with a higher transmission coefficient.

Panels (c)-(f) of Figure \ref{fig:T_coeff_Bifrost}
present the power spectral density distributions of the 
synthetic observables and the actual plasma vertical velocity. In each panel, we
plot the PSD of the vertical plasma velocity at the formation height 
of the spectral line (v$_{sim}$) as well as the PSD of the Doppler velocity 
measured in the synthetic observables for solar region of interest.
The lines are the mean of the 
PSD distributions and the shaded areas represent the 10th to 90th percentiles region 
of the distributions. For the \ion{Mn}{1} line on average the power
spectra of the observed Doppler velocity signals is attenuated with a constant shift, as 
seen previously in Paper I and discussed in previous work
\citep{1980A&A....84...96M, 2009A&A...508..941B}.
However, for the plage-like
regions the power spectra of the true plasma velocity
and the one of the observed (synthetic) Doppler velocity
are very similar and in some places the true velocity power exceeds the
observed one. This is due to the fact that the passing
wave fronts in the atmosphere introduce a jump-like 
change of the height of formation of the diagnostics, introducing
jump-like signals in the measured Doppler velocity. This effect cannot be 
described with 1D semiempirical atmospheric modeling and we believe
that it is important to be included in the accurate estimates of the wave fluxes,
as it will increase the attenuation coefficient significantly for brighter regions, leading
to lower acoustic flux estimates.

As is evident, the attenuation coefficient varies significantly and may 
depend in part on the solar feature being observed. 
This effect cannot be captured by static 
1D models and will be definitely misrepresented by 1D hydrodynamic 
time-dependent models, as the obvious dependence on the simulated magnetic 
topology of the solar region helps determine its value. 
Hence, we believe that future estimations of the acoustic flux in the 
chromosphere should take the complicated nature of the transmission 
coefficient into consideration.

\section{Inferring the acoustic wave flux}
\label{sec:Inferred_flux}

\begin{table}[!ht]
	\begin{center}

		\begin{tabular}{l|c|c} 
			\textbf{Spectral line} & Density $\rho$ [kg m$^{-3}$] &  $\cal{T}$ \\ \hline
			QS \ion{Na}{1} D$_1$ 589.6 nm & 6.55 10$^{-7}$ & 0.68 \\
			QS \ion{Mn}{1} 280.1 nm & 2.96 10$^{-8}$ &  0.37\\
			QS \ion{Ca}{2} 854.2 nm & 3.15 10$^{-9}$ & 0.56 \\
			QS \ion{Mg}{2} k3 & 8.55 10$^{-12}$ & 0.50\\
		    Plage \ion{Na}{1} D$_1$ 589.6 nm & 2.04 10$^{-6}$ & 0.91 \\
			Plage \ion{Mn}{1} 280.1 nm & 5.82 10$^{-7}$&  1.03\\
			Plage \ion{Ca}{2} 854.2 nm & 1.02 10$^{-8}$ & 0.89 \\
			Plage \ion{Mg}{2} k3 & 5.53 10$^{-11}$& 0.80\\					
		\end{tabular}
		\caption{Density and attenuation coefficient values
			used for the acoustic flux estimation derived from the averages 
		    of the corresponding representative regions in 
		    Figure~\ref{fig:Bifrost_overview}. 
	    }
		
	\end{center}
	\label{tab:Flux_quantities}
\end{table}

\begin{figure}[!htbp]
	\includegraphics[width=0.45\textwidth]{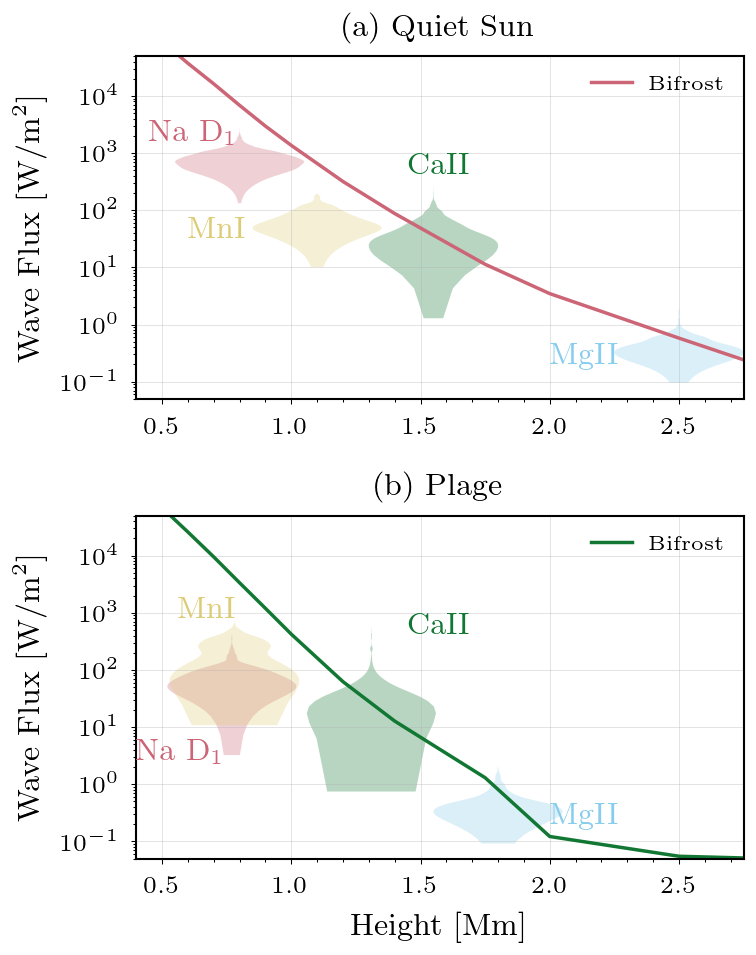}
	\caption{Acoustic flux in the different solar regions 
		inferred from the IRIS observations presented in
		Figure~\ref{fig:Vrms_total_power}.  Panel (a) (top)
		presents the fluxes inferred for the internetwork region
		and the bottom panel (b) presents the fluxes 
        inferred for the plage region. The straight lines show the acoustic fluxes in the 
	    Bifrost simulation at the respective height for the solar feature.}
\label{fig:Wave_flux}
\end{figure}

Based on the observational data presented in 
Section~\ref{sec:Observations} and the numerical analysis in 
Sections~\ref{sec:Modeling} and \ref{sec:Uncertainties},
we have the required physical quantities
to demonstrate the method for estimating the acoustic flux 
from the IRIS and IBIS observations, based on the 
synthetic observables derived from the Bifrost models.
In Section~\ref{sec:Uncertainties} we showed that the internetwork
and the plage regions exhibit different line formation 
characteristics, such as densities, velocity formation regions and attenuation of 
the wave signals, within the Bifrost simulation. 
In particular, the 
internetwork exhibits formation of the line 
that is significantly lower in density and 
has a lower transmission coefficient, compared
to the active network elements.

Despite the variances in density shown in Figure~\ref{fig:rho_comparison}
we chose to use the densities from Bifrost to compute acoustic wave 
fluxes. The mean formation properties employed for the different 
spectral lines are listed in Table~\ref{tab:Flux_quantities} and were derived
from averaging over the representative regions (shown as small boxes)
in Figures~\ref{fig:Bifrost_overview} and \ref{fig:T_coeff_Bifrost}.
The attenuation coefficient values we obtained are significantly closer to unity
than what previous authors have cited 
\citep{2009A&A...508..941B,2020ApJ...890...22A}, which could be due to the 
different (and more realistic) modeling approach we employed. However, this will also lead
to significantly lower estimates of the energy fluxes.

 We used the average formation properties, as derived for regions of the simulation 
 for the observed internetwork and plage regions described in Section~\ref{sec:Uncertainties}. 
To calculate the wave fluxes, we adopt values of the density and 
the attenuation coefficient for the spectral lines separately for the two 
regions (shown as squares 
in Figure~\ref{fig:Bifrost_overview}). The values are listed in
Table~\ref{tab:Flux_quantities}.

Figure~\ref{fig:Wave_flux} presents the estimated wave 
fluxes, based on the calculated properties of the Bifrost 
simulations for the corresponding
solar features. The top panel (a) shows the diagnostics for the internetwork
and the bottom panel (b) shows the results for the plage regions. Overplotted 
are the Bifrost averaged acoustic flux as a function of height 
for the two regions for comparison. 

For the case of the internetwork, the observation-derived values were generally 
lower than the acoustic fluxes retrieved directly from the simulations. This can 
be traced back to the difference in the observed and simulated Doppler 
velocity distributions in Figure~\ref{fig:Vrms_total_power}.
This might be due to a variety of reasons, including the magnetic
field topology, incorrect driving of the p-modes in the bottom 
boundary of the simulations \citep{2021RSPTA.37900170F}
or incomplete physical treatment of the wave propagation and dissipation.

For the case of the plage observations, we saw that the 
lower chromosphere diagnostics were orders of magnitudes below 
the fluxes present in the simulation. Our 
modeling approach shows that the two independently observed lines of 
\ion{Na}{1} D$_1$ and the \ion{Mn}{1} 280.1 nm exhibit almost
 the same amount of acoustic flux at the about same formation height, acting as a
 self-consistency check. 
The wave fluxes derived from the middle and upper chromospheric diagnostics 
exhibit values closer to the ones derived from the Bifrost models.

In conclusion, our analysis shows that the observed Doppler velocities are lower than 
what were derived from synthetic line profiles calculated from simulations in the four spectral 
lines with formation heights spanning the chromosphere. 
Hence, the acoustic fluxes calculated from the observations,
based on the simulation results, are also 
lower and likely insufficient to maintain the solar chromosphere in its quiescent state.
This conclusion holds for both the quiet and plage chromospheres, 
compared to their respective
radiative losses \citep{1976ASSL...53.....A}.
However, our analysis shows that any such conclusions are highly model dependent.
In particular, the biggest systematic biases are the 
estimates of the (average-value) densities and the attenuation coefficients.

\section{Discussion and conclusions}
\label{sec:Conclusions}

We present UV observations 
of waves in the solar chromosphere with the IRIS spacecraft.
In particular, we examined the 
spectral lines of \ion{Mn}{1} 280.1 nm (upper photosphere/lower chromosphere)
and the \ion{Mg}{2} k3 feature (upper chromosphere).  
Reduction steps, described in Section~\ref{sec:Observations}
enhance the data and the wave signatures were readily observed. 
The power spectra of the observed Doppler velocities
and line-core intensity oscillations described in Section~\ref{sec:power_spectra}
exhibit the previously seen 
ubiquitous power law distributions. Comparing them with optical
diagnostics from previous studies in the literature, we find agreement between
the oscillatory properties of UV and optical diagnostics separately observed 
with IRIS and IBIS.

To interpret these observations, we relied on the 3D rMHD simulation 
Bifrost, that provided us with a detailed model of the lower solar atmosphere.
This model includes detailed physics (non-LTE radiative losses and dynamic
hydrogen ionization) important
for wave propagation. We used the synthetic observables from 
\cite{2013ApJ...778..143P} complemented with our own RH15D synthesis
to understand the formation of the spectral diagnostics in question,
described in Section~\ref{sec:Modeling}.
We found that the average density and heights of 
formation of the spectral lines differ significantly 
between the internetwork and network regions.
Therefore, the height corresponding to the plasma velocity sampled by the Doppler
measurement in these lines also changes with the underlying solar feature.

We compared the formation properties of the discussed spectral 
lines with other wave-modeling approaches used in the literature -- the
RADYN code and 1D semiempirical atmospheric perturbative
approaches. In Section~\ref{sec:Uncertainties}, we examined the 
differing  formation properties resulting from the different modeling
approaches and how they affect the inferred fluxes. In particular, we 
discuss how the measured Doppler velocities correspond to 
actual atmospheric velocities at different height for the different solar 
features in Section~\ref{subsec:uncertainty_vel_fluc}. In Bifrost we saw a strong
a notable separation in
the density of formation for the internetwork and the plage regions. 
The density of formation is significantly lower than the values
found in previous work
based on 1D semiempirical models. The value of the
transmission coefficient is also significantly lower for the
internetwork than for the enhanced network regions, too. 
However, it is significantly higher than values used in previous work, leading to a 
lower acoustic wave flux estimates.

Finally, in Section~\ref{sec:Inferred_flux} we presented the inferred wave 
fluxes based on the physical parameters derived from the Bifrost simulations
shown in Table~\ref{tab:Flux_quantities}. We used
the values for internetwork and enhanced network separately. 
In our analysis the wave fluxes inferred from the observations
are lower than the ones found in the simulation. In particular,
the acoustic fluxes in the lower solar atmosphere, around the formation height
of \ion{Mn}{1} and Na D$_1$ lines are about a few hundred W/m$^2$. At the formation
 heights of the \ion{Mg}{2} k3 feature, they are on the order of a few
 W/m$^2$. These results do not disagree per se with previous ones in the
 literature, more than what would be expected due to the systematic modeling biases
described in Section~\ref{sec:Uncertainties}. 
 
Our work provides us with an example how more realistic simulations of the solar
atmosphere are important for understanding the solar and stellar chromospheres.
In particular, we show that the
observed velocity field is not directly related to a singular height
in the solar atmosphere. As shown in Figure~\ref{fig:Bifrost_v_rms_vs_flux}, 
there is no good correlation between observed velocity amplitudes and the actual
wave flux at the height of formation of the line.
Because the velocity fluctuations  
are our key observable for atmospheric energetics, this means 
that our abilities to derive the amount of acoustic flux is severely limited. We show that 
the density of formation and transmission coefficients have 
to be adopted for different solar regions to be able to infer the acoustic flux. 

However, given the complex structuring of the chromosphere, 
we will remain dependent on 3D rMHD models and the derived synthetic observables to 
provide the basis for determining certain statistical characteristics of different regions of 
the complex atmosphere. Yet, this makes our derivation of values like the acoustic flux 
dependent on the veracity and accuracy of those models, which in itself is a challenge 
to accurately ascertain. It is necessary to be aware of the uncertainties and systematic biases 
carried forward by values based on these models.


\acknowledgements 

\vspace{1mm}
\facilities{IRIS, DST(IBIS).}

\software {SolarSoft; Matplotlib \citep{matplotlib}; NumPy \citep{numpy}; 
SciPy \citep{scipy}; h5Py; RH15D \citep{RH_15D}.
The Python and IDL scripts 
utilized for this project are available 
on the public repository of the author: 
\url{https://github.com/momomolnar/IRIS_wave_signatures}}. 

\acknowledgements

IRIS is a NASA
small explorer mission developed and operated by LMSAL with
mission operations executed at NASA Ames Research center and
major contributions to downlink communications funded by ESA
and the Norwegian Space Centre.
Data in this publication were obtained with
the facilities of the National Solar Observatory, which is operated 
by the Association of Universities for Research in Astronomy, 
Inc. (AURA), under cooperative agreement 
with the National Science Foundation. 
The authors would like to thank 
the anonymous referee, Amanda Alexander, Karen Slater, and Rahul Yadav whose 
input greatly improved the manuscript.
MEM was supported for part of this work by the 
DKIST Ambassador Program, funding for which is provided by the National Solar
Observatory, a facility of the National Science Foundation, operated under Cooperative 
Support Agreement number AST-1400405; and in part by a FINESST fellowship with grant
number 80NSSC20K1505. This work utilized resources from
 the University of Colorado Boulder Research Computing Group, 
 which is supported by the National Science Foundation
(awards ACI-1532235 and ACI-1532236), the University of
Colorado Boulder, and Colorado State University. 

\bibliography{Bibliography.bib}{}



\end{document}